\documentclass[showpacs,twocolumn,superscriptaddress,nofootinbib]{revtex4}

\usepackage{doi}
\usepackage{hyperref}
\hypersetup{
  colorlinks=true,        
  linkcolor=blue,         
  citecolor=cyan,         
}

\usepackage{graphicx}
\usepackage{dcolumn}
\usepackage{bm}
\usepackage{color}
\usepackage{footmisc}

\usepackage{amsmath}
\usepackage{amssymb}


\begin{document}

\title{Testing the weak cosmic censorship conjecture for a Reissner-Nordstr\"{o}m-de Sitter black hole surrounded by perfect fluid dark matter}

\author{Sanjar Shaymatov}
\email{sanjar@astrin.uz}

\affiliation{Institute for Theoretical Physics and Cosmology, Zheijiang University of Technology, Hangzhou 310023, China}\affiliation{Akfa University, Kichik Halqa Yuli Street 17,  Tashkent 100095, Uzbekistan}
\affiliation{Ulugh Beg Astronomical Institute, Astronomicheskaya 33, Tashkent 100052, Uzbekistan}
\affiliation{National University of Uzbekistan, Tashkent 100174, Uzbekistan}
\affiliation{Tashkent Institute of Irrigation and Agricultural Mechanization Engineers,\\ Kori Niyoziy 39, Tashkent 100000, Uzbekistan}

\author{Bobomurat Ahmedov} 
\email{ahmedov@astrin.uz}

\affiliation{Ulugh Beg Astronomical Institute, Astronomicheskaya 33, Tashkent 100052, Uzbekistan} 
\affiliation{National University of Uzbekistan, Tashkent 100174, Uzbekistan}
\affiliation{Tashkent Institute of Irrigation and Agricultural Mechanization Engineers,\\ Kori Niyoziy 39, Tashkent 100000, Uzbekistan}

\author{Mubasher Jamil}
\email{mjamil@zjut.edu.cn}

\affiliation{Institute for Theoretical Physics and Cosmology, Zheijiang University of Technology, Hangzhou 310023, China}
\affiliation{Department of Mathematics, School of Natural Sciences (SNS), National
University of Sciences and Technology (NUST), H-12, Islamabad 44000, Pakistan}
\affiliation{Canadian Quantum Research Center 204-3002 32 Ave Vernon, BC V1T 2L7, Canada}

\date{\today}
\begin{abstract}
In this paper, we test the weak cosmic censorship conjecture (WCCC) for the Reissner-Nordstr\"{o}m-de Sitter (RN-dS) black hole surrounded by perfect fluid dark matter. We consider a spherically symmetric perturbation on deriving linear and non-linear order perturbation inequalities by applying new version of gedanken experiments well accepted from the work of Sorce and Wald. Contrary to the well-known result that the Reissner-Nordstr\"{o}m (RN) black hole could be overcharged under linear order particle accretion it is hereby shown that the same black hole in perfect fluid dark matter with cosmological parameter cannot be overcharged. Considering a realistic scenario in which black holes can not be considered to be in vacuum we investigate the contribution of dark matter and cosmological constant in the overcharging process of an electrically charged black hole. We demonstrate that the black hole can be overcharged only when two fields induced by dark matter and cosmological parameter are completely balanced. Further we give a remarkable result that black hole cannot be overcharged beyond a certain threshold limit for which the effect arising from the cosmological constant dominates over the effect by the perfect fluid dark matter. Thus even for linear accretion process, the black hole cannot always be overcharged and hence obeys the WCCC in general. This result would continues be fulfilled for non-linear order accretion. 

\end{abstract}
 \maketitle

\section{Introduction}
\label{introduction}

In General Relativity (GR), astrophysical black holes are formed under gravitational collapse of massive star at the end state of evolution and have always been very fascinating and intriguing objects for their extreme geometric and remarkable gravitational properties. The existence of black holes has been predicted by simple mathematical models as a generic result of finding exact analytical solution of the field equations of GR. Recent gravitational wave signals detected by the LIGO and Virgo scientific collaborations \cite{Abbott16a,Abbott16b} and the first image of supermassive black hole at the center of the elliptical M87 galaxy, observed by the Event Horizon Telescope (EHT) Collaboration~\cite{Akiyama19L1,Akiyama19L6} have provided the strong evidence in favor of existence of black holes in nature. Those observations have been expected to be very potent tests in probing unknown aspects associated with precise measurements of the parameters of black holes, yet there still remains open questions concerning the behaviour of black holes. We know that Einstein's gravity --General Relativity, is a best described theory in the strong field regime apart from the singular regions of the spacetime. However, GR is an incomplete theory due to the inevitable occurrence of singularity which has remained one of the most important unresolved questions~\cite{Hawking-Penrose70}. Theoretical existence of singularity marks the limit of GR where it loses its applicability. In this respect, the cosmic censorship conjecture (CCC) was first proposed by Penrose~\cite{Penrose69} in 1969 for imposing the validity of the Einstein gravity, i.e. the WCCC always prevents the singularity from being seen by outside observers.  However, the WCCC still remains open as there exists no strong proof for its validity in the general case. Despite this fact, the WCCC has been tested in the context of different gravity models via proposed gedanken experiments to understand whether or not a black hole turns into a naked singularity. If it were possible to violate the WCCC, it would lead to serious implications from observational point of view, i.e. it would make it possible that one can observe the final state of gravitational collapse of an object. Besides there does exist a vast literature on the formation of naked singularity as an end state of gravitational collapse, starting from the Christodoulou (1986)~\cite[see,
e.g.][]{Christodoulou86,Joshi93,Joshi00,Goswami06,Harada02,Stuchlik12a,
Vieira14,Stuchlik14,Giacomazzo-Rezzolla11,Joshi15}].

The validity of the WCCC was formulated by Wald~\cite{Wald74b} for the first time to overcharge/overspin black hole by the process of particle accretion. In such a process it was shown that turning an extremal black hole into naked singularity can never happen by using test particles with appropriate parameters and the WCCC is strongly held. Later this issue was addressed somewhat differently \cite{Dadhich97} and it was shown that it is impossible for particles of suitable parameters to approach horizon as parameter space pinches off. Thus, a nearly extremal black hole cannot be turned into the extremal one by falling in particles. Later on it was found \cite{Hubeny99} that extremality may however be jumped over in a discontinuous process and hence the horizon can be destroyed. It means that a black hole could be overcharged by particle accretion process. Much later this experiment was extended to Kerr and Kerr-Newman black holes \cite{Jacobson09,Saa11}. It was shown that a rotating black hole could be overspun if and only if falling in particle adds a sufficient amount of angular momentum to black hole's angular momentum. Sufficient amount of work has been done since then to test the WCCC for overcharging/overspinning of black hole in  various  frameworks, \cite[see, e.g.][]{Matsas07,Shaymatov15,Bouhmadi-Lopez10,Rocha14,Jana18,Song18,Duztas18,Duztas-Jamil18b,Duztas-Jamil20,Yang20a,Yang20b}. It is worth noting that in the above extensive body of works, the higher order and backreaction effects were ignored. Whereas if those effects are taken into account it would not be possible all through for impinging particles to destroy black hole horizon and thus the WCCC is always respected~\cite[see,e.g.][]{Barausse10,Zimmerman13,Rocha11,Isoyama11,
Colleoni15a,Colleoni15b,Li13}. Furthermore, the above thought experiment has been extended to the context of magnetized black holes~\cite{Siahaan16,Shaymatov19b}, black hole with charged scalar field \cite{Gwak20}, BTZ black holes \cite{Duztas16} as well as black hole dynamics~\cite{Mishra19,Bardeen73b}. Further, a clear distinction between black hole and naked singularity was shown through the phenomenon of spin precession~\cite{Chakraborty17}. 

Recently Sorce and Wald \cite{Sorce-Wald17,Wald18} have developed a new version of gedanken experiment, which allows one to take non-linear order perturbation process. It was shown that this experiment strongly supports the validity of the WCCC, i.e.  black hole cannot be overcharged/spun under non-linear order perturbations. It turns out that this thought experiment can only provide the correct result whether black hole could be overcharged/spun. In this context, the WCCC cannot be violated all through under non-linear order~\cite{An18,Gwak18a,Ge18,Ning19,Yan-Li19,Shaymatov19a,Shaymatov19b,Jiang20plb}. Further the same analysis has been extended to the higher dimensional black holes. Note that the WCCC has already been explored for a higher dimensional charged black hole, leading to the result that black hole could be overcharged under linear order perturbation~\cite{Revelar-Vega17}. It is then worth studying a charged rotating black hole in five dimensions whether it can be over-extremalized.  In spite of the fact that Kerr-Newman black hole has no exact solution of Einstein-Maxwell equation for an analogue of a five dimensional Kerr-Newman black hole this issue can be fixed by considering the minimally gauged supergravity charged rotating black hole in five dimensions~\cite{Chong05}.  It is shown that when one goes to five dimensional charged rotating black hole with single rotation, it could be  over-extremalized in case when charge parameter dominates over rotation while it cannot do so when the opposite is the case~\cite{Shaymatov19c}. It is worth noting that the study of overspining of an extremal black hole in higher dimensions shows that neither an extremal black hole could be overspun nor non-extremal one be converted into extremal under linear order perturbation~\cite{Bouhmadi-Lopez10}. Further the WCCC has been addressed for black holes in dimensions $D\geq 5$~\cite{Shaymatov19a,Shaymatov20a}, and the black hole in dimensions $D\geq 6$ can never be overspun and always obeys the CCC in the weak form~\cite{Shaymatov21a}.   

In an astrophysical context it is believed that black holes can not be found in vacuum due to the presence of matter and fields in their nearby environment. Cosmological observations of supernova explosions SNIa confirm an accelerating rate of expansion of our universe at present, commonly explained by a cosmological constant $\Lambda$ endowed with a repulsive gravitational effect (i.e. de Sitter case). The cosmological observations also suggest the estimated value of the cosmological constant to be $\Lambda\sim10^{-52} m^{-2}$ \cite{Peebles03,Spergel07}. Thus, taking into account the repulsive effect due to the cosmological constant would play an important role at large length and time scales as well as in the black hole vicinity. It is a fact that the motion of test particles can be drastically influenced by the geometry in the strong field regime while at the same time both the geometry and geodesics of such particles could be affected by other matter fields as well. With this respect, the existence of dark matter is particularly important similarly to the cosmological constant being important at large scales.  The idea of existence of dark matter has initially been introduced by observation of the flat rotation curves of giant elliptical and spiral galaxies~\cite{Rubin80}. It is believed in the light of mounting evidence from the astrophysical data that the rotational velocity of stars in the outskirts of many giant spiral galaxies can only be explained with the help of elusive dark matter which contributes to approximately up to 90 \% mass of the galaxy while the rest is the luminous matter composed of baryonic matter \cite{Persic96}. In the early phase of evolution of the universe, the dark matter used to be mostly found near the galactic central regions which helped in the formation and clustering of stars around the galactic center while in the late stages of galactic evolution, the dark matter gradually drifted far out to form a dark matter galactic halo around the host galaxy by various dynamical processes. Dark matter has not been so far detected directly, yet astrophysical observations indicate that many giant elliptical and spiral galaxies contain a (sometimes a binary) supermassive black hole in the galactic center embedded in a giant dark matter halo \cite{Akiyama19L1,Akiyama19L6,Nayak12,Haroon20,Jusufi20,Hendi20}. In literature, several black hole solutions with a dark matter background have been proposed (see for example \cite{Kiselev03-dm,Li-Yang12,Xu18,Xu16-dm,Hou18-dm,Haroon19,Rizwan19,Jusufi19,Konoplya19plb,Narzilloev20b,Shaymatov20b}). {The analysis of the accretion process for different spherically symmetric space-time geometries for a static fluid and the accretion of phantom energy onto a stationary charged black hole has been investigated in testing various conditions~(see for example \cite{Bahamonde15,Jamil08})}. 

We note that the above analysis has also been extended to the case of rotating anti-de Sitter (AdS) black holes~\cite{Gwak16,Natario16,Natario20} as well as to the asymptotically AdS case~\cite{Zhang14,Gwak16JCAP,Gwak16PLB} to check the validity of the WCCC. It is also worth noting that the WCCC has also been recently addressed \cite{Wang20} for a RN-AdS black hole where cosmological constant is regarded as a variable. It is well known \cite{Hubeny99} that the WCCC for RN black hole can be violated at the linear order accretion and this was also proved by new version of gedanken experiment \cite{Sorce-Wald17} for linear order and the result is however overturned when non-linear order process was included. The question then arises, what happens for RN black hole when we consider real astrophysical scenario by taking into account effects due to the presence of cosmological constant and perfect fluid dark matter--could it be overcharged or not and does it violate the WCCC for linear order accretion?  This is what we plan to investigate in this paper.  

The paper is organized as follows: In Sec.~\ref{Sec:IW} we review and describe variational identities for a diffeomorphism covariant theory to derive linear and non-linear variational identities. In Sec.~\ref{Sec:EM-RN} we briefly discuss Einstein-Maxwell theory and the metric for the RN-dS black hole in the perfect fluid dark matter. In Sec.~\ref{Sec:GN} we derive perturbation inequalities for linear and non-linear order perturbations and check whether overcharging of black hole is possible or not. We end up with conclusion in Sec.~\ref{Sec:Conclusion} which summarises the obtained results.

\section{Variational identities }\label{Sec:IW}

To derive variational identities, a diffeomorphism covariant theory  was proposed for manifold $\mathcal M$ in $n$- dimensions. This theory stems from a Lagrangian $\mathbf{L}$ composed of metric $g_{ab}$ and other fields $\psi$~\cite{Wald94,Sorce-Wald17}. One can then indicate all dynamical fields by $\phi=(g_{ab},\psi)$. Thus, the above Lagrangian is defined by  
\begin{eqnarray}\label{Eq:Lagrangian}
\delta \mathbf{L}=\mathbf{E} \delta\phi+d\mathbf{\Theta}(\phi,\delta\phi)\, ,
\end{eqnarray}
with $\mathbf{E}$ being the equations of motion, and ${\mathbf\Theta}$ being the symplectic potential. The symplectic current $(n-1)$-form $\mathbf{\omega}$ then takes the form 
\begin{eqnarray}
\mathbf{\omega}(\phi,\delta_1\phi,\delta_2\phi)=\delta_1 \mathbf{\Theta}(\phi,\delta_2\phi)-\delta_2\mathbf{\Theta}(\phi,\delta_1\phi)\, .
\end{eqnarray}
One may consider an arbitrary vector $\xi^{a}$ with $\phi$ for given $M$ in order to define the Noether current $(n-1)$-form 
\begin{eqnarray}\label{Eq:NR-charge}
\mathbf{J}_\xi=\mathbf{\Theta}(\phi,{L}_\xi\phi)-\xi \cdot\mathbf{L}\, .
\end{eqnarray}
{Following to \cite{Wald95} the above equation for current can be written as}
\begin{eqnarray}\label{Eq:NR-charge1}
\mathbf{J}_\xi=d\mathbf{Q}_\xi+ \mathbf{C}_\xi\, ,
\end{eqnarray}
with $\mathbf{Q}_\xi$ being the Noether charge. The second term   $ \textbf{C}_\xi=\xi^{a}\textbf{C}_{a}$ on the right hand side in the above expression is referred to as the constraint of the theory and  vanishes in the case when the equations of motion are satisfied, i.e. for $d\mathbf{J}_\xi=0$.

Further Eqs. (\ref{Eq:NR-charge}) and (\ref{Eq:NR-charge1}) give rise to the condition for obtaining the linear variational identity for fixed $\xi^{a}$, and thus on a Cauchy surface $\Xi$ we have 
\begin{eqnarray}\label{Eq:first-order}
\int_{\partial\Xi}\left[\delta \mathbf{Q}_\xi-\xi\cdot\mathbf{\Theta}(\phi,\delta\phi)\right]&=&\int_{\Xi}\mathbf{\omega}(\phi,\delta\phi,\mathcal{L}_\xi\phi)\nonumber\\&-&\int_{\Xi}\xi\cdot\mathbf{E}\delta\phi-\int_{\Xi}\delta \mathbf{C}_\xi \, .
\end{eqnarray}
The first term $\int_{\Xi}{\omega}(\phi,\delta\phi,\mathcal{L}_\xi\phi)$ in the above equation corresponds to the variation part of the system for a given vector field $\xi^{a}$. This term vanishes as long as $\xi^{a}$ satisfies a Killing vector and a symmetry of $\phi$. {Since $\xi^{a}$ is considered to be a Killing vector being invariable in the perturbation process the following is satisfied, i.e. $\mathcal{L}_\xi\phi=0$.} Considering the linear variational identity, we obtain the non-linear variational identity on the same surface  
\begin{eqnarray}\label{Eq:second-order}
\int_{\partial\Xi}\left[\delta^2 \mathbf{Q}_\xi-\xi\cdot\delta\mathbf{\Theta}(\phi,\delta\phi)\right]&=&\int_{\Xi}{\omega}(\phi,\delta\phi,\mathcal{L}_\xi\delta\phi)\nonumber\\ &-&\int_{\Xi}\xi\cdot\delta\mathbf{E}\delta\phi\nonumber\\&-&\int_{\Xi}\delta^2 \mathbf{C}_\xi\, .
\end{eqnarray}

\section{Einstein-Maxwell theory and RN-dS black hole in perfect fluid dark matter}\label{Sec:EM-RN}

{In this section, we consider Einstein-Maxwell theory for obtaining an explicit form for linear and non-linear variational identities for nearly extremal RN-dS black hole in perfect fluid dark matter by adapting the well accepted method developed by Sorce and Wald~\cite{Sorce-Wald17}. However it turns out that the dark matter profile contained in the Lagrangian may lead to more complicated variational identities. To avoid this we now consider the off-shell variation of Einstein-Maxwell theory~\cite{Jiang20}. The Lagrangian in four dimensional Einstein-Maxwell theory is given by
\begin{eqnarray}\label{Eq:EM-Lag}
\mathbf{L}=\frac{\mathbf{\epsilon}}{16\pi}\left(R-F^{a b}F_{a b}\right)\, ,
\end{eqnarray}
with the volume element $\mathbf\epsilon$ for given black hole spacetime  metric, the Ricci scalar $R$ and Faraday tensor of electromagnetic field $F_{ab}$. Let us then define the dynamical field which is composed of the metric and vector potential of electromagnetic field, i.e. $\phi=(g_{ab},A_{a})$ satisfying the above Lagrangian and we have    
\begin{eqnarray}\label{Eq:E}
\mathbf{E}(\phi)\delta\phi=-{\epsilon}\left(\frac{1}{2} T^{ab}
\delta g_{a b}+\textbf{j}^a\delta \textbf{A}_{a}\right)\, ,
\end{eqnarray}
with the non-electromagnetic part of the stress-energy tensor 
\begin{eqnarray}
T_{ab}=\frac{1}{8\pi}\left(R_{ab}-\frac{1}{2}g_{ab}R\right)-T_{ab}^{EM}\, , 
\end{eqnarray} and the electromagnetic current $\textbf{j}^a=\frac{1}{4\pi} \nabla_{b} F^{ab}$. Note that the corresponding non-electromagnetic part consists of two parts, ordinary mater and dark matter sources, i.e. $T_{ab}=T^{M}_{ab}+T^{DM}_{ab}$.} From the above equation, the symplectic potential consists of two electromagnetic and gravity parts and can be defined in the following way 
\begin{eqnarray}\label{Eq:symplectic}
\mathbf{\Theta}_{ijk}\left(\phi,\delta\phi\right)&=&\frac{1}{16\pi}\epsilon_{aijk}g^{ab} g^{cd}(\nabla_{d}\delta g_{bc}-\nabla_{b}\delta g_{cd})\nonumber\\&-&\frac{1}{4\pi}\epsilon_{aijk}F^{ab}\delta \textbf{A}_{b}\, , 
\end{eqnarray}
where $\epsilon_{aijk}$ is Levi-Civita tensor. 
For the Einstein-Maxwell theory the corresponding symplectic current takes the following form   
 \begin{eqnarray}\label{Eq:sym-current}
\omega_{ijk}&=&\frac{1}{4\pi}\left[\delta_{2}(\epsilon_{aijk} F^{ab}) \delta_{1} \textbf{A}_{b}-\delta_{1}(\epsilon_{aijk} F^{ab}) \delta_{2} \textbf{A}_{b}\right]\nonumber\\&+&\frac{1}{16\pi}\epsilon_{aijk} w^{a}\, ,
\end{eqnarray}
where the first term on the right hand side corresponds to the electromagnetic part while the second -- gravity part.  From Eq.~(\ref{Eq:sym-current}) $w^{i}$ is given by 
\begin{eqnarray}
w^{i}&=&P^{ijkhab}\left(\delta_{2} g_{jk} \nabla_{h}\delta_{1}g_{ab}-\delta_{1} g_{jk} \nabla_{h}\delta_{2}g_{ab}\right)\, ,
\end{eqnarray}
with
\begin{eqnarray} 
P^{ijkhab}&=&g^{ia} g^{b j} g^{k h}-\frac{1}{2}g^{ih} g^{ja} g^{b k} - \frac{1}{2}g^{i j} g^{k h} g^{ab}\nonumber\\ &-& \frac{1}{2}g^{j k} g^{ia} g^{b h} + \frac{1}{2}g^{jk} g^{i h} g^{ab}\, .
 	\end{eqnarray}
By employing $\mathcal{L}_{\xi} g_{ab} = \nabla_{a}\xi_{b} +\nabla_{b}\xi_{a}$ and $\nabla_{a}\textbf{A}_{b}=F_{ab}+\nabla_{b}\textbf{A}_{a}$, we obtain the Noether current  
\begin{eqnarray}\label{Eq:NC}
(J_{\xi})_{ijk}&=&\frac{1}{8\pi}\epsilon_{aijk } \nabla_{b}(\nabla^{[b} \xi^{a]}) + \epsilon_{aijk} T_{b}^{a} \xi^{b}\nonumber\\&+&\frac{1}{4\pi}\epsilon_{aijk}\nabla_{c}(F^{ca} \textbf{A}_{b} \xi^{b}) + \epsilon_{aijk} \textbf{A}_{b} \textbf{j}^{a} \xi^{b}\, . 
\end{eqnarray}
From Eq.~(\ref{Eq:NR-charge1}), the Noether charge $Q_{\xi}=Q_{\xi}^{GR}+Q_{\xi}^{EM}$ can be written in the following form 
\begin{eqnarray}
(Q_{\xi})_{ijk}&=&-\frac{1}{16\pi}\epsilon_{ijk ab}\nabla^{a}\xi^{b}-\frac{1}{8\pi}\epsilon_{ijk ab} F^{ab } \textbf{A}_{c} \xi^{c}\, ,
\end{eqnarray}
and the constraint 
\begin{eqnarray}
(C_{\zeta})_{ijk}&=&\epsilon_{aijk }(T_{\zeta}^{a} + \textbf{A}_{\zeta} \textbf{j}^{a})\, . 
\end{eqnarray}  

Next, we consider a static and spherically symmetric charged black hole spacetime metric as a solution of Einstein-Maxwell theory. The RN black hole metric generalized to a nonvanishing cosmological parameter $\Lambda$ has been introduced by Lake \cite{Lake79}, a metric known as the RN-dS metric. Also there was a solution that provides a way to include a dark mater distribution in black hole background geometry~\cite{Li-Yang12}. Later this solution was generalised to incorporate nonvanishing cosmological constant \cite{Xu18}. {Here, we consider spacetime metric which describes a static and spherically symmetric RN-dS black hole in perfect fluid dark matter for which the Lagrangian of Einstein-Maxwell theory is given by
\begin{eqnarray}\label{Eq:EM-Lag}
\mathbf{L}=\frac{\mathbf{\epsilon}}{16\pi}\left(R-2\Lambda-F^{a b}F_{a b}+16\pi\mathcal{L}_{DM}\right)\, ,
\end{eqnarray}
with cosmological constant $\Lambda$ with positive value, and the dark matter Lagrangian density $\mathcal{L}_{DM}$. According to the above Lagrangian the Einstein equation of motion can be written 
\begin{eqnarray}
R_{ab}-\frac{1}{2}g_{ab}R+\Lambda g_{ab}&=&8\pi\left(T_{ab}^{DM}+T_{ab}^{EM}\right)\, , \\
 \nabla_{b} F^{ab}&=&{4\pi} \textbf{j}^a\, , 
\end{eqnarray}
with $T^{EM}_{ab}$ representing the stress energy tensor for the electromagnetic field  
\begin{eqnarray}
T^{EM}_{ab}=\frac{1}{4\pi}\left(F_{ac}F^c_{b}-\frac{1}{4}g_{ab}F_{cd}F^{cd}\right)\, ,  
\end{eqnarray}
while $T_{ab}^{DM}$ for the perfect fluid dark matter and specified as 
\begin{eqnarray}
(T^{a}_{b})^{DM}={\rm diag}(-\rho,p_{r},p_{\theta},p_{\phi})\, .  
\end{eqnarray}
In the above equation, $\rho$, $p_{r}$, $p_{\theta}$ and $p_{\phi}$ respectively refer to the density, radial and tangential pressures. Then, a static and spherically symmetric RN-dS black hole metric in perfect fluid dark matter is given by~\cite{Li-Yang12,Xu16-dm}  
}
\begin{eqnarray} \label{Eq:metric}
d s^2 &=& \, -F(r) dt^2 + F(r)^{-1}\, dr^2  +  r^2d\Omega^2\, ,
  \end{eqnarray}
with line element of 2-sphere $d\Omega^2$ and
\begin{eqnarray}
 F(r)&=& 1- \frac{2M}{r} + \frac{Q^2}{r^{2}}-\frac{\Lambda}{3}r^2+\frac{\lambda}{r}\log\frac{r}{\vert\lambda\vert}\, . 
 \end{eqnarray}
where parameters $M$ and $Q$ are the mass and electric charge of the black hole while $\Lambda$ and $\lambda$ are related to the cosmological constant and the perfect fluid dark matter, respectively. The metric (\ref{Eq:metric}) reduces to the RNdS black hole in the case of vanishing $\lambda=0$, while it reduces to the RN black hole in case both parameters $\lambda$ and $\Lambda$ tend to zero. 
 
The corresponding electromagnetic potential is given by 
\begin{eqnarray} \label{el:pot} \textbf{A} = -\frac{Q}{r} dt \, . \end{eqnarray}

Note that here we shall focus on the positive $\Lambda>0$ and $\lambda>0$ in testing the effects of cosmological constant and dark matter distribution upon the overcharging of black hole.  For black hole horizon, one can write the following relation 
\begin{eqnarray}\label{Eq:hor}
\frac{\Lambda}{3}r^4 -r^2+ {2M r} -r{\lambda}\log\frac{r}{\vert\lambda\vert}- {Q^2}=0\, ,
\end{eqnarray}
which has three positive real roots, i.e. the inner horizon $r_{-}$, the outer horizon $r_{+}$ and the cosmological horizon $r_{c}$ being the largest one among those three roots. However, we shall restrict ourselves to the first two roots, inner and outer horizons. In the case of small values of $\Lambda M^2\ll 1$ and $\lambda/M\ll 1$, Eq.~(\ref{Eq:hor}) gives the approximated analytical form for black hole horizon as
\begin{eqnarray} \label{Eq:hor1}
 r_{\pm}=M \pm \sqrt{M^2-Q^2+\frac{\Lambda}{3} M^4- \lambda M \log \frac{M}{\vert\lambda \vert}}\, .
 \end{eqnarray}
From the the above expression for black hole horizon it is immediately clear that 
the second term 
\begin{eqnarray}\label{Eq:f}
f=M^2-Q^2+\frac{\Lambda}{3} M^4- \lambda M \log \frac{M}{\vert\lambda \vert}\geq 0\, ,
\end{eqnarray}
must be always positive for the existence of black hole horizon, while if it is negative definite it turns into a naked singularity. When this term goes to zero, the two horizons then coincide, corresponding to the near extremal black hole. Thus, the black hole could be overcharged, similarly to the RN black hole that can be overcharged. Further we study whether the overcharging holds well or not.  

{At the given horizon $r_{+}$, the horizon area, the surface gravity, the electromagnetic potential and new physical quantity for RN black hole surrounded by dark matter with cosmological parameter will be respectively written as follows \cite{Xu16-dm}: 
\begin{eqnarray}\label{A}
A_{+}&=& 4\pi r_{+}^2\, ,\\
\label{k}
k &=& \frac{f'}{2}|_{r=r_+}\, ,\\
\label{Phi}
\Phi_{+}&=& \frac{Q}{r_{+}}\, ,\\
\label{Pi}
\Pi_{+} &=& \frac{1}{2} \log\frac{r_{+}}{\vert\lambda\vert}\, .
\end{eqnarray}  
From the above parameters we can write the first law of black hole dynamics 
\begin{eqnarray}
\delta M-\Phi_{+}\delta Q-\Pi_{+}\delta\lambda = \frac{k}{8\pi}\delta A\, ,
\end{eqnarray}
where the surface gravity $k$ tends to zero for an extremal black hole. }
 
\section{Perturbation inequalities and gedanken experiment to overcharge a RN-dS black hole in perfect fluid dark matter  }\label{Sec:GN}

It is known that the falling of test particle into the black hole transfers the mass and charge to its mass and charge. Hence the final state of the black hole parameters is then defined by $M+\delta M$, $Q+\delta Q$ and $\lambda+\delta \lambda$, respectively. For final state it was shown by Hubeny \cite{Hubeny99} that a nearly extremal RN black can be overcharged to $M<Q$ state by linear order particle accretion. In this section we approach this issue by considering new gedanken experiment proposed by Sorce and Wald \cite{Sorce-Wald17}. {This experiment allows one to include a non-linear order perturbation process, according to which a one-parameter family of field $\phi(\alpha)$ perturbation in the background spacetime is taken into consideration. For the sake of clarity, $\phi(0)$ is referred to as the black hole solution, as described by the line element shown in Eq.~(\ref{Eq:metric}). Here we consider that $\phi(\alpha\neq 0)$ exhibits a dynamic spherically symmetric solution perturbed by falling in matter field, and thus the equation of motion can be written as   
\begin{eqnarray}
R_{ab}(\alpha)&-&\frac{1}{2}R(\alpha)g_{ab}(\alpha)+\Lambda g_{ab}(\alpha)\nonumber\\&=&8\pi\left[T_{ab}^{M}(\alpha)+T_{ab}^{DM}(\alpha)+T_{ab}^{EM}(\alpha)\right]\, , \\
\nabla_{b} F^{ab}(\alpha)&=&4\pi \textbf{j}^a(\alpha)\, .
\end{eqnarray}}
{We note that $\phi(0)$ satisfies a stationary RN-dS black hole solution with perfect fluid dark matter background, i.e., $T_{ab}^{M}(0)=\textbf{j}^{a}(0)=0$.  
For this one-parameter family of field $\phi(\alpha)$, all matter sources are assumed to cross the horizon portion and fall into the black hole. For that, we apply the Eddington-Finkelstein coordinate $\upsilon=t+\int dt/F(r)$ to obtain the following line element  
\begin{eqnarray}
d s^2(\alpha) &=& \, -F(r,\upsilon,\alpha) d\upsilon^2 + 2\mu(r,\upsilon,\alpha)\,d\upsilon dr  +  r^2d\Omega^2\, ,\nonumber\\
\end{eqnarray}
which, at the same time, describes the dynamical geometry after being perturbed by matter field. Note that 
$$
F(r,\upsilon,0)=F(r)=1- \frac{2M}{r} + \frac{Q^2}{r^{2}}-\frac{\Lambda}{3}r^2+\frac{\lambda}{r}\log\frac{r}{\vert\lambda\vert}
$$
and $\mu\left(r,\upsilon,\alpha=0,1\right)=1$ satisfy the background geometry in the above spacetime (see for example~\cite{Jiang20b}).     
Here we assume the spacetime geometry we are considering is linearly stable to the one-parameter family of field perturbation $\phi(\alpha)$, thus referring to the stability condition implying that the spacetime geometry approaches another one described by $M(\alpha)$, $Q(\alpha)$ and $\lambda(\alpha)$ at the sufficiently late times under the spherically symmetric perturbation arising from the falling in matter source, i.e. the dynamical fields can be described by 
\begin{eqnarray} 
F(r,\upsilon,\alpha)=F(r,\alpha)\,\,\,   \mbox{and}\,\,\, \textbf{A} = -\frac{Q(\alpha)}{r} d\upsilon\, .
\end{eqnarray} } 
{For this family of perturbation we have already chosen a hypersurface $\Xi=\Xi_{1}\cup H$. $\Xi$ defines a region which starts from the bifurcation surface $B$ at one end and extends to the horizon portion $H$, so it becomes spacelike $\Xi_{1}$ at the other end, and tends to the asymptotic flatness at the infinity. The above hypersurface $\Xi=\Xi_{1}\cup H$ was shown by diagram in detail in Ref.~\cite{Sorce-Wald17}. Here we note that we assume the spacetime geometry satisfies the above stability condition stating that the perturbation vanishes at the bifurcation surface $B$ from the property of hypersurface $\Xi$.} 

Further we study the variational identities for a nearly extremal black hole. Let us first consider the linear variational identity by the perturbation caused by falling in mater. Since the first term on the right hand side vanishes accordingly for a Killing vector $\xi^{\alpha}$ satisfying $\mathcal{L}_\xi\phi=0$, the linear variational identity on the hypersurface $\Xi$ in the expression~(\ref{Eq:first-order}) takes the following form 
%
%
%
\begin{eqnarray}\label{Eq:eq-motion}
\int_{\partial\Xi}\left[\delta \mathbf{Q}_\zeta-\zeta\cdot\mathbf{\Theta}(\phi,\delta\phi)\right]&=&-\int_{\Xi}\zeta\cdot\mathbf{E}\delta\phi\nonumber\\&&-\int_{\Xi}\delta \mathbf{C}_\zeta \, .
\end{eqnarray}
where $\zeta^{a}$ is assumed to be vector field for exterior solution of a stationary black hole. Then one can consider $\zeta^{a}=\zeta_{(t)}^{a}$ as the timelike Killing vector, satisfying both equations of motion, i.e. $\mathcal{L}_\zeta\phi=0$. Hereafter, we focus only on the vector field  $\zeta^{a}$, and  {for further procedure we choose the gauge condition satisfying 
\begin{eqnarray}
\zeta^{a}\delta \textbf{A}(\alpha)_{a}|_{r=r_+}=0\, ,
\end{eqnarray}
where $\zeta^{a}=\left(\partial/\partial\upsilon\right)^{a}$ and $r_{+}$ respectively refer to the timelike Killing vector field and the horizon radius.}

{With the above procedure we consider the boundaries of Cauchy surface $\Xi=\Xi_{1}\cup H$ which consists of two boundaries interpreted by  spatial infinity at one end and the bifurcation surface $B$ at the other. Here we adapt the properties of the hypersurface and spacetime metric as well, and hence the left-hand side of Eq.~(\ref{Eq:eq-motion}) can be separated in the following form
\begin{eqnarray}\label{Eq:surfaces}
\int_{\partial\Xi}\left[\delta \mathbf{Q}_\zeta -\zeta\cdot\mathbf{\Theta}(\phi,\delta\phi)\right]&=&\int_{S_\infty}\left[\delta \mathbf{Q}_\zeta-\zeta\cdot\mathbf{\Theta}(\phi,\delta\phi)\right]\nonumber\\&-&\int_{B}\left[\delta \mathbf{Q}_\zeta-\zeta\cdot\mathbf{\Theta}(\phi,\delta\phi)\right]\, .\nonumber\\ 
\end{eqnarray}
{The point to be noted here is that we apply $S_{\infty}$ sphere to replace the boundary of the surface $\Xi_{1}$. Hence,  the contribution to the boundary integral $S_{\infty}$  stems from the gravitational part~\cite{Sorce-Wald17,Jiang20} and is given by 
\begin{eqnarray}\label{Eq:infinity}
\int_{S_{\infty}}\left[\delta \mathbf{Q}_\zeta-\zeta\cdot\mathbf{\Theta}(\phi,\delta\phi)\right]&=&\delta M
\, .
\end{eqnarray}}
By using Eqs.~(\ref{Eq:eq-motion}) and (\ref{Eq:surfaces}), we rewrite Eq.~(\ref{Eq:first-order}) for the linear variational identity 
{\begin{eqnarray}\label{Eq:first-order1}
\delta M
&=&\int_B[\delta \mathbf{Q}_\zeta-\zeta\cdot\mathbf{\Theta}(\phi,\delta\phi)]-\int_{\Xi_{1}}\zeta\cdot\mathbf{E}\delta\phi\nonumber\\&&-\int_{\Xi_{1}}\delta \mathbf{C}_\zeta-\int_H\delta \mathbf{C}_\zeta\, .
\end{eqnarray}
Note that the first term on the right hand side vanishes at the bifurcation surface from the properties of the hypersurface $\Xi$.  Employing Eq.~(\ref{Eq:E}) we derive the second term of Eq.~(\ref{Eq:first-order1}) as    
\begin{eqnarray}
\int_{\Xi_{1}}\zeta\cdot\mathbf{E}\delta\phi = -\int_{\Xi_{1}}\zeta\cdot{\epsilon}\left(\frac{1}{2} T^{ab}
\delta g_{a b}+\textbf{j}^a\delta \textbf{A}_{a}\right)\, .
\end{eqnarray}
Here, we assume that the the perturbation satisfies the stability condition, i.e., $T_{ab}^{M}(\alpha)=\textbf{j}^{a}(\alpha)=0$ at the $\Xi_1$ surface~\cite{Jiang20}. By imposing this condition the total stress-energy tensor is given by $T_{ab}(\alpha)=T_{ab}(\alpha)^{DM}$ for the background spacetime. Thus one can have 
\begin{eqnarray}\label{Eq:st}
T^{ab}(\alpha)\frac{d g_{ab}(\alpha)}{d\alpha}=0\, ,
\end{eqnarray}      
which leads to the second term on the right hand side of Eq.~(\ref{Eq:first-order1}) to vanish at the $\Xi_1$. The third term can be obtained as    
\begin{eqnarray}
\int_{\Xi_{1}}\delta \mathbf{C}_\zeta=- \Pi_{+} \delta \lambda\, ,
\end{eqnarray}
where we have used $[C_{\zeta}(\alpha)]_{ijk}=\epsilon_{aijk }[T_{h}^{a}(\alpha)\zeta^{h} + \textbf{A}_{h}(\alpha) \textbf{j}^{a}(\alpha)\zeta^{h}]$ with the volume element $\tilde{\epsilon}_{ijk}$.  
The rest term on the right hand side of Eq.~(\ref{Eq:first-order1}) then stems from
\begin{eqnarray} \label{Eq:zeta1} 
\int_H\delta \mathbf{C}_\zeta &=&\int_{H}\epsilon_{aijk }\zeta_{(t)}^{b}\Big(\delta T_{b}^{a} + \textbf{A}_{b} \delta \textbf{j}^{a}\Big)\, .
 \end{eqnarray}
Taking $\Phi_{+}=-\zeta^{b}\textbf{A}_{b}\vert_{H}$ into consideration with $\int_{H}\delta(\epsilon_{aijk}\textbf{j}^{a})=\delta Q$ and the gauge condition $\zeta^{b} \delta \textbf{A}_{b}\vert_{H}=0$, Eq.~(\ref{Eq:first-order1}) yields 
\begin{eqnarray}
\delta M-\Phi_{+}\delta
Q-\Pi_{+}\delta\lambda = - \int_{H}\epsilon_{aijk} \zeta_{b} \delta T^{ab}\, .
\end{eqnarray}
For the volume element one may write $\epsilon_{aijk}=-4k_{[a} \tilde{\epsilon}_{ijk]}$ on the horizon portion $H$.  With this we consider only the case for which all matter fields satisfy {the null energy condition, i.e., $\delta T_{ab}(\alpha)k^{a}(\alpha)k^{b}(\alpha)\geq 0$ always for any null vector $k^{a}(\alpha)$~\cite{Sorce-Wald17,Jiang20}.}} Thus, the linear order variational inequality can be written in the following form
\begin{eqnarray}
\label{Eq:first-order2}  
\delta M-\Phi_{+}\delta Q-\Pi_{+}\delta\lambda\geq 0\, .
\end{eqnarray}

{Following the linear order inequality we further obtain the non-linear order variational identity for a near extremal black hole. Similarly, Eq.~(\ref{Eq:second-order}) for the non-linear variational identity is given by
{\begin{eqnarray}\label{Eq:second-order1}
\int_{S_\infty}[\delta^2 \mathbf{Q}_\zeta &-&\zeta\cdot\delta\mathbf{\Theta}(\phi,\delta\phi)]=\int_B[\delta^2 \mathbf{Q}_\zeta-\zeta\cdot\delta\mathbf{\Theta}(\phi,\delta\phi)]\nonumber\\&-&\int_{\Xi_1}\delta\left(\zeta\cdot\mathbf{E}\delta\phi\right) -\int_{H}\delta\left(\zeta\cdot\mathbf{E}\delta\phi\right) \nonumber\\&-&\int_{\Xi_1}\delta^2 \mathbf{C}_\zeta-\int_{H}\delta^2 \mathbf{C}_\zeta +\mathcal{E}_\Xi(\phi,\delta\phi)\, ,
\end{eqnarray}
where for the left-hand side we can obtain 
\begin{eqnarray}
\int_{S_\infty}[\delta^2 \mathbf{Q}_\zeta -\zeta\cdot\delta\mathbf{\Theta}(\phi,\delta\phi)]=\delta^2 M\, .
\end{eqnarray}
Note that $\mathcal{E}_\Xi(\phi,\delta\phi)$ in Eq.~(\ref{Eq:second-order1}) the above equation refers to the canonical energy and is defined by the non-linear perturbation $\delta\phi$ on $\Xi$. For the second term on the right hand side we have 
\begin{eqnarray}
\int_{\Xi_1}\delta\left(\zeta\cdot\mathbf{E}\delta\phi\right) = -\frac{1}{2} \int_{\Xi_{1}}\zeta\cdot{\epsilon}\,\delta\left[T^{ab}(\alpha)\delta g_{ab}(\alpha)\right]\, . 
\end{eqnarray}
By imposing the stability condition at the $\Xi_{1}$ surface and Eq.~(\ref{Eq:st}) the above integral can easily reduce to zero. Since the killing vector $\Xi^{a}$ is tangent to the horizon portion $H$ the third term on the right hand side in Eq.~(\ref{Eq:second-order1}) can be neglected on $H$.  For the fourth term we obtain 
\begin{eqnarray}
\int_{\Xi_{1}}\delta^2 \mathbf{C}_\zeta=- \Pi_{+} \delta^2 \lambda\, , 
\end{eqnarray} 
 while the fifth term can be written as 
\begin{eqnarray}
\int_{\Xi}\delta^2 \mathbf{C}_\zeta &=&\int_{H}\epsilon_{aijk}\zeta_{(t)}^{b}\left(\delta^2 T_{b}^{a} + \textbf{A}_{b} \delta^2 \textbf{j}^{a}\right)\nonumber\\&=& \int_{H}\tilde{\epsilon}_{ijk} k_{a}\zeta_{b} \delta^2 T^{ab}+\Phi_{+}\delta^2
Q\, , 
\end{eqnarray}
where we applied the gauge condition $\zeta^{a}\delta \textbf{A}_{a}=0$ on $H$ of $\Xi=\Xi_{1}\cup H$ with $\zeta^{a}$ being tangent to $H$. As before we substitute the null energy condition $\delta^2 T_{ab}k^{a}k^{b}\geq0$ in the above equation. For non-linear order variational identity,  taking into consideration above results, we rewrite Eq.~(\ref{Eq:second-order1}) as  
\begin{eqnarray}\label{Eq:second-order2}
\delta^2 M-\Phi_{+}\delta^2
Q-\Pi_{+}\delta^2 \lambda &=& \int_B[\delta^2 \mathbf{Q}_\zeta-\zeta\cdot\delta\mathbf{\Theta}(\phi,\delta\phi)]\nonumber\\&+&\mathcal{E}_{\Xi_1}(\phi,\delta\phi)+\mathcal{E}_{H}(\phi,\delta\phi)\, . \nonumber\\
\end{eqnarray}  
}
Further we define a one-parameter field perturbation as $\phi(\alpha)^{RN-dS}$ induced by falling in matter absorbed by the RN-dS black hole surrounded by perfect fluid dark matter with following parameters 
\begin{eqnarray}\label{MQ}
M(\alpha)&=& M+\alpha\delta M\, ,\nonumber\\
Q(\alpha)&=&Q+\alpha\delta Q\, , \nonumber\\
\lambda(\alpha)&=&\lambda+\alpha\delta \lambda\, ,
\end{eqnarray}
with $\delta M$, $\delta Q$ and $\delta \lambda$ chosen to satisfy the linear order perturbation given by Eq.~(\ref{Eq:first-order2}). {We may then evaluate the rest of the terms of Eq.~(\ref{Eq:second-order2}) for $\phi^{RN-dS}$. 
Since $\delta^2 M=\delta^2 Q_{B}=\delta^2 \lambda_{B}=\mathcal{E}_{H}(\phi,\delta\phi^{RN-dS})=0$ and as well as $\mathcal{E}(\phi,\delta\phi^{RN-dS})$ can be neglected on $\Xi_1$ for this family we have} 
\begin{eqnarray}
\delta^2 M&-&\Phi_{+}\delta^2
Q-\Pi_{+}\delta^2 \lambda\nonumber\\& =&\int_B[\delta^2 \mathbf{Q}_\zeta-\zeta\cdot\delta\mathbf{\Theta}(\phi,\delta\phi^{RN-dS})]\, .
\end{eqnarray}
From the property of $\Xi$ the vector $\zeta^{a}$ vanishes at the bifurcation surface $B$, i.e. $\zeta^{a}=0$, so we come to the non-linear variational identity as      
\begin{eqnarray}\label{Eq:non-linear}
\delta^2 M-\Phi_{+}\delta^2 Q-\Pi_{+}\delta^2 \lambda\geq -\frac{k}{8\pi}\delta^2 A^{RN-dS}\, .
\end{eqnarray}

Following this new version of gedanken experiment based on the procedure described above we study overcharging of 
a nearly extremal RN-dS black hole in perfect fluid dark matter. Now we recall Eq.~(\ref{Eq:f}) for which extremality is indicated by $f=0$ while existing horizon by $f>0$. As was mentioned above for RN black hole $f<0$ leads to the destruction of its horizon. To test whether this condition really happens or not for the RN black hole in the perfect fluid dark matter with cosmological constant, i.e. $M^2-Q^2+\frac{\Lambda}{3} M^4- \lambda M \log \frac{M}{\vert\lambda \vert}< 0$, we apply new version of gedanken experiment \cite{Sorce-Wald17} that allows to consider one parameter family of perturbation function $f(\alpha)$ leading to the inclusion of higher order perturbations. Finally one can write  
{\begin{eqnarray}\label{Eq:par1}
f(\alpha)&=&M(\alpha)^2-Q(\alpha)^2+\frac{\Lambda}{3} M(\alpha)^4 \nonumber\\ &-&\lambda(\alpha) M(\alpha) \log \frac{M(\alpha)}{\vert\lambda(\alpha) \vert}\, ,
\end{eqnarray}
with $M(\alpha)$, $Q(\alpha)$  and $\lambda(\alpha)$} given by Eq.~(\ref{MQ}). It is worth noticing that Eq.~(\ref{Eq:par1}) reduces to the RN case when $\Lambda M^3/3=\lambda \log M/\vert \lambda\vert$. It means that the two fields induced by perfect fluid dark matter and cosmological constant are completely balanced, their contribution will not be included all through for overcharging of black hole. This then leads to an interesting question -- could those two fields contribute to the overcharging of black hole? The only way to settle this question is to consider the general case in which the ratio of two fields is written as 
\begin{eqnarray}\label{Eq:ratio}
\Lambda = \frac{3\beta}{M^3}\lambda\log \frac{M}{\vert \lambda\vert}\, ,
\end{eqnarray}
where $\beta \geq 1 $, $\leq 1$, respectively. In $\beta > 1 $ the field due to cosmological parameter dominates over the field as that of perfect fluid dark matter, while the latter does for $\beta < 1 $. Therefore we rewrite Eq.~(\ref{Eq:par1}) as
{\begin{eqnarray}\label{Eq:par2}
f(\alpha)&=&M(\alpha)^2-Q(\alpha)^2+ \lambda(\alpha) (\beta-1)M(\alpha) \log \frac{M(\alpha)}{\vert\lambda(\alpha) \vert}\, .\nonumber\\
\end{eqnarray}}
From Eq.~(\ref{Eq:par2}) $f(0)=M^2\epsilon^2$ refers to a near extremal black hole with $\epsilon \ll 1$, while for $\alpha \neq 0$ we expand the function $f(\alpha)$ up to second order in $\epsilon$ and $\alpha$ as  
%
 \begin{eqnarray}\label{Eq:par3}
f(\alpha)=M^4\epsilon^2+f_1\alpha+f_2\alpha^2+O(\alpha^3, \alpha^2\epsilon,\alpha\epsilon^2,\epsilon^3)\, ,
\end{eqnarray}
where $f_1$ and $f_2$ respectively refer to 
linear and non-linear perturbations. 
Eq.~(\ref{Eq:par3}) clearly shows that $f(\lambda)<0$ allows transition from black hole to naked singularity, thereby overcharging can be attained. Thus we further show whether that is really attainable or not. Let us then explore the linear $f_1$ and non-linear $f_2$ perturbations which are given by  
{\begin{eqnarray}\label{Eq:f1}
f_{1}&=&\left[2 M+\lambda  (\beta-1)  \left(1+\log \frac{M}{\vert\lambda\vert }\right)\right]\delta M -2 Q \delta Q\nonumber\\
&-&(\beta-1)M\left(1-\log \frac{M}{\vert\lambda\vert }\right)\delta\lambda\, , 
\end{eqnarray}
and
\begin{eqnarray}\label{Eq:f2}
f_{2}&=& \left[M+\frac{\lambda  (\beta -1)}{2}\left(1+ \log \frac{M}{\vert\lambda \vert}\right)\right]\delta^2 M -Q \delta^2 Q\nonumber\\&-&
\frac{M}{2}(\beta-1)\left(1-\log \frac{M}{\vert\lambda \vert}\right)\delta^2\lambda\nonumber\\
&+&\left(1+\frac{\lambda  (\beta-1) }{2 M}\right)\delta M^2 -\delta Q^2\nonumber\\&+&(\beta-1)\left(\log \frac{M}{\vert\lambda \vert}\delta M-\frac{M}{2\lambda}\delta\lambda\right)\delta \lambda\, .
\end{eqnarray}}
Next we intend to define $\delta M$ being the minimum possible value required for overcharging of black hole \cite{Hubeny99,Sorce-Wald17}. So $\delta M$ is defined by       
{\begin{eqnarray}\label{Eq:min}
\delta M_{min} & \geq &\frac{Q}{r_{+}}\delta Q +\frac{1}{2}\log \frac{r_{+}}{\vert\lambda \vert}\delta \lambda=\frac{Q }{M}\delta Q \left(1-\epsilon\right) \nonumber\\&+&\frac{1}{2}\left(\log \frac{M}{\vert\lambda \vert}-\log (1-\epsilon)\right)\delta\lambda+ O(\epsilon^2)\, .
\end{eqnarray}}
This is the minimal possible energy required for test particles to cross the horizon and falling into the black hole. 

Bearing in mind $\delta M$ we first rewrite $f(\alpha)$ for linear order perturbation  
{\begin{eqnarray}\label{Eq:linear1}
f(\alpha)&=& M^2\epsilon^2 + \left[2 M+\lambda  (\beta-1)  \left(1+\log \frac{M}{\vert\lambda\vert }\right)\right]\nonumber\\&\times &\left\{\delta M - \left(\frac{Q}{M}\delta Q+\frac{\beta-1}{2}\left(1-\log \frac{M}{\vert\lambda\vert }\right)\delta\lambda\right) \right. \nonumber\\&\times & \left.\left[1 -\frac{\lambda}{2M}  (\beta-1) \left(1+\log \frac{M}{\vert\lambda\vert }\right)\right] \right\}\alpha +\mathcal O(\alpha^2)\, .\nonumber\\
\end{eqnarray}}
Let us first consider $\beta =1$ referring to the case in which two fields induced by perfect fluid dark matter and cosmological constant are completely balanced as seen from Eq.~(\ref{Eq:ratio}). Hence, the above equation takes the form 
\begin{eqnarray}\label{Eq:RN1}
f(\alpha)&=& M^2\epsilon^2 - 2 Q ~\delta Q ~\epsilon \alpha +\mathcal O(\alpha^2)\, .
\end{eqnarray}
This clearly shows that it is possible to make $f(\alpha)<0$; thus, it turns out that the RN-dS black hole surrounded by perfect fluid dark matter can be overcharged under linear order perturbation in the case of $\beta=1$. Also Eq.~(\ref{Eq:RN1}) corresponds to the result of RN black hole case obtained by Source and Wald \cite{Sorce-Wald17}. Let us then come to the $f(\alpha)$ to explore it for general case, i.e. $\beta\neq 1$. Taking into account Eq.~(\ref{Eq:min}) for charged test particle we rewrite $f(\alpha)$
{\begin{eqnarray}\label{Eq:linear2}
f(\alpha)&=& M^2\epsilon^2 + \left[2 M+\lambda  (\beta-1)  \left(1+\log \frac{M}{\vert\lambda\vert }\right)\right]\nonumber\\&\times &\left\{\frac{Q}{2M^2} \lambda (\beta-1) \left(1+\log \frac{M}{\vert\lambda\vert }\right)\delta Q \right.\nonumber\\ &+& \frac{1}{2}\left[1-\beta\left(1-\log \frac{M}{\vert\lambda\vert }\right)\right]\delta \lambda\nonumber\\&-&\left. \left(\frac{Q}{M}\delta Q-\frac{\delta \lambda}{2}\right)\epsilon\right\}\alpha +\mathcal O(\alpha^2)\, .
\end{eqnarray}}
In order to make $f(\alpha)>0$ the second term in the above equation must always be positive definite, providing that the following inequality is satisfied:  
{\begin{eqnarray}\label{Eq:limit1}
\beta \geq 1+\frac{  2 M Q \epsilon \delta Q  - M^2 \left(\epsilon  +\log \frac{M}{\vert\lambda \vert}\right)\delta\lambda }{Q \lambda  \delta Q -M^2\delta \lambda  +\Big( M^2\delta \lambda +\lambda  Q\delta Q \Big)\log \frac{M}{\vert \lambda \vert} } \, .
\end{eqnarray}
Note that $\delta Q>\delta \lambda$ always holds good. In the case of $\delta Q\gg\delta \lambda$, Eq.~(\ref{Eq:limit1}) reduces to 
\begin{eqnarray}\label{Eq:limit2}
\beta \geq 1+\frac{2M\epsilon}{\lambda \left(1+\log \frac{M}{\vert\lambda \vert} \right)} \, .
\end{eqnarray}}
Eq.~(\ref{Eq:limit1}) indicates the threshold limit beyond which black hole cannot be overcharged and hence would always obey the WCCC even under linear order accretion. Let us consider the numerical example for parameter $\beta$: Setting $M=1$ we choose $\lambda=0.001$ with $\epsilon=0.01$, so we have $\beta=3.52916$ (For this thought experiment one can use different values of $\lambda$ and $\epsilon$; see Fig.~\ref{fig}). With this, we have shown the critical value for $\beta$ beyond which relative dominance of cosmological parameter over dark matter parameter occurs. Hence, the field required due to the cosmological parameter would be slightly stronger as compared to the one due to perfect fluid dark matter.
\begin{figure}
\centering
  \includegraphics[width=0.45\textwidth]{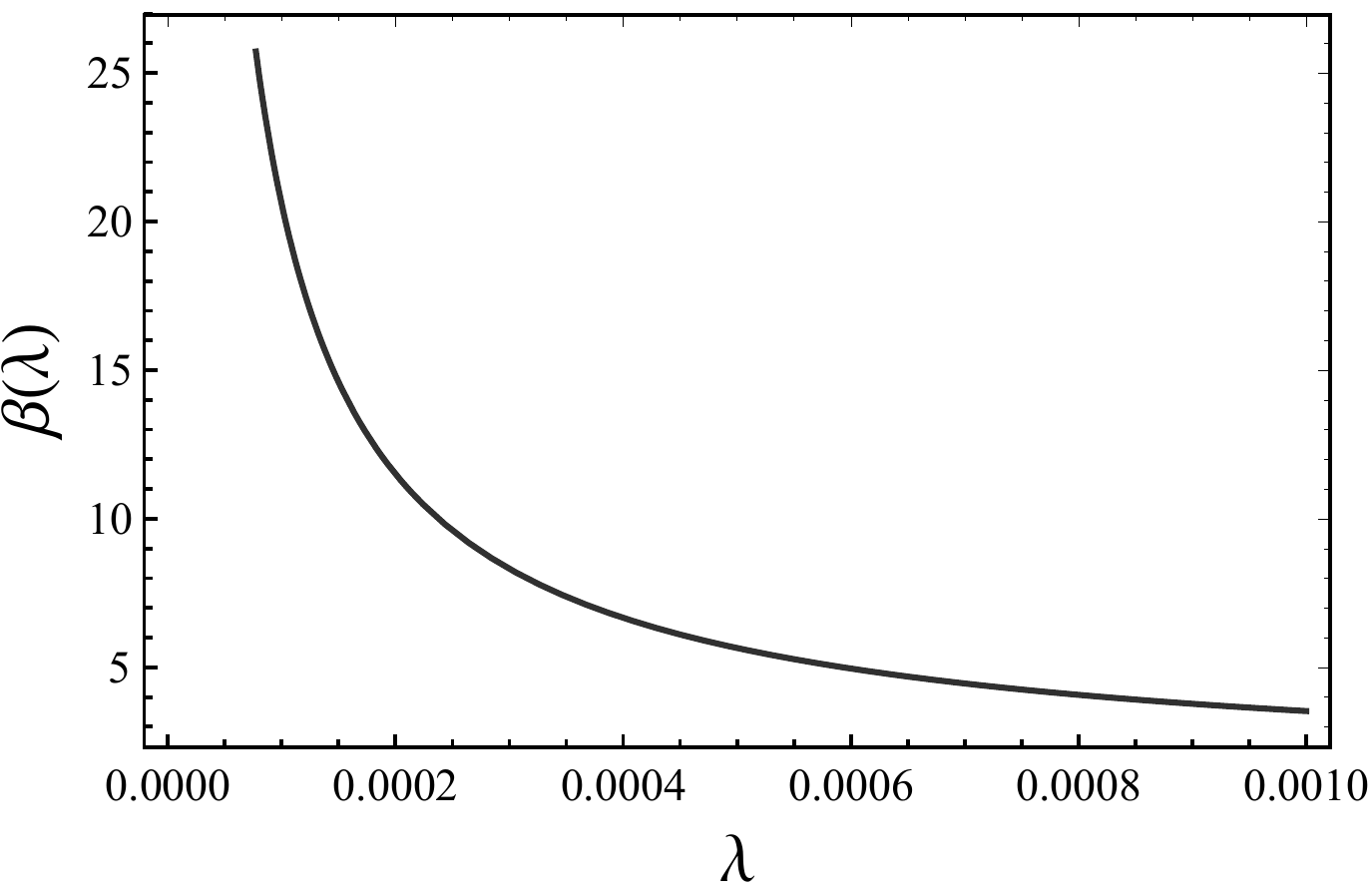}

\caption{\label{fig} {The plot illustrates the values of threshold parameter $\beta$ as a function of the perfect fluid dark matter parameter $\lambda$ in the case of fixed $\epsilon=0.01$. Note that beyond the threshold value of $\beta$ black hole always favours no overcharging even under linear order accretion.} }
\end{figure} 

Now we consider a non-linear order perturbation and recall Eq~(\ref{Eq:f2})  
{\begin{eqnarray}\label{Eq:f22}
f_{2}&\approx & \left[M+\frac{\lambda  (\beta -1)}{2}\left(1+ \log \frac{M}{\vert\lambda \vert}\right)\right]\nonumber\\&\times &\Big(\delta^2 M -\Phi_{+}\delta^2 Q-\Pi_{+}\delta^2 \lambda \Big)\nonumber\\&+&\frac{M}{2}\left(1+\beta\log \frac{M}{\vert\lambda \vert}-\beta\right)\delta^2\lambda-\delta Q^2\nonumber\\
&+&\left(1+\frac{\lambda  (\beta-1) }{2 M}\right)\delta M^2 + (\beta-1)\log \frac{M}{\vert\lambda \vert}\delta M\delta\lambda\, ,\nonumber\\
\end{eqnarray}}
where non-linear terms are defined by following inequality 
{\begin{eqnarray}
\delta^2 M -\Phi_{+}\delta^2 Q-\Pi_{+}\delta^2 \lambda\geq -\frac{k}{8\pi}\delta^2 A\, .
\end{eqnarray}}
Recalling Eqs.~(\ref{A}) and (\ref{k}) and employing Eq.~(\ref{Eq:min})  
we can rewrite Eq.~(\ref{Eq:f22}) as
{\begin{eqnarray}
f_{2} &\approx &  \frac{1}{4M^2}\left[M+\frac{\lambda  (\beta -1)}{2}\left(1+ \log \frac{M}{\vert\lambda \vert}\right)\right]\nonumber\\&\times &\bigg[ \Big(M \left(4 \delta Q^2+\delta \lambda ^2\right)+4 (\beta -2) Q\delta Q \delta \lambda  \Big)\nonumber\\&+&\log \frac{M}{\vert\lambda \vert} \bigg(2 \Big((\beta-1) M \delta \lambda-4 \beta Q\delta Q \Big)\delta \lambda\nonumber\\&+&  \frac{\beta-1}{M}  \lambda  \Big( M\delta \lambda -10  Q\delta Q\Big)\delta \lambda  \nonumber\\&-&\log \frac{M}{\vert\lambda \vert} \Big( M(4 \beta -1) \delta \lambda ^2+(\beta-1)  \lambda  \log \frac{M}{\vert\lambda \vert}\delta \lambda ^2\nonumber\\ &+&  \frac{(\beta-1)}{2M}  \lambda  (8 Q\delta Q  +3 M \delta \lambda )\delta \lambda  \Big)\bigg)\nonumber\\&-& 2 \frac{(\beta-1)}{M^2}  \lambda  Q \Big( Q\delta Q+2 M \delta \lambda \Big)\delta Q  \bigg]\nonumber\\&+&\left(1+\frac{\lambda  (\beta-1) }{2 M}\right)\delta M^2-\delta Q^2\nonumber\\&+& (\beta-1)\log \frac{M}{\vert\lambda \vert}\delta M\delta\lambda\, ,
\end{eqnarray}
and for $\delta Q \gg \delta \lambda$ it can be approximated by the following analytical expression
\begin{eqnarray}
f_{2}&\approx & \left[M+\frac{\lambda  (\beta -1)}{2}\left(1+ \log \frac{M}{\vert\lambda \vert}\right)\right]\nonumber\\ &\times & \left[ M^4-\frac{\lambda\left(\beta -1\right)}{2}   M Q^2 \right]\frac{\delta Q^2}{M^5}\nonumber\\
&+&\left[1+\frac{\lambda  (\beta-1) }{2 M}\right]\delta M^2 -\delta Q^2\, .
\end{eqnarray}
From the above equations one can see that $f_2\geq0$ is always satisfied since $\delta M\geq \Phi_{+}\delta Q+\Pi_{+}\delta \lambda$ following from Eq.~(\ref{Eq:first-order2}) continues to hold good for test particle.} Hence, we have $f(\alpha)\geq 0$, and no overcharging occurs. That means since $f(\alpha)\geq0$ always, thus the transition from black hole to over-extremal state can never happen. As expected the black hole 
always favours no overcharging when non-linear order perturbations are involved, and thereby the WCCC is always respected. {It is worth noting that in a realistic scenario the values of perfect fluid dark matter and cosmological constant are very small and thus the obtained conclusions hold under the assumptions $\Lambda,\lambda \ll 1$.}      
   

\section{Conclusions}
\label{Sec:Conclusion}

{In this paper, we studied the validity of the WCCC for the RN-dS black hole with perfect fluid dark matter profile by adapting new version of the gedanken experiments developed by Sorce and Wald. We derived perturbation inequalities for linear and non-linear order perturbations under the spherically symmetric perturbation. } It is well known \cite{Hubeny99} that the RN black hole can be overcharged and thus the WCCC is violated for linear accretion process while it would always be restored when non-linear order perturbations are included. It is believed that in a realistic scenario astrophysical objects can not be considered to be in vacuum due to the cosmological constant and the existence of dark matter distribution surrounding objects. The question is, could RN black hole be overcharged at the linear order accretion when one considers the real astrophysical scenario as that of effects due to the cosmological constant and perfect fluid dark matter? This is what we have addressed in this paper. As we know the cosmological constant ($\Lambda>0$) is defined by a repulsive behaviour of space expansion while dark matter by attractive one. We have shown that for linear accretion the RN-dS black hole surrounded by perfect fluid dark matter can be overcharged when two fields due to the dark matter and cosmological constant are completely balanced. However, it is overturned when we include non-linear order perturbations and thus the question of its overcharging never arise. Then further analysis led to the remarkable result. We have shown that black hole cannot be overcharged beyond a certain threshold limit for which repulsive effect arising from the cosmological constant dominates over the attractive one due to the perfect fluid dark matter. Though even for linear accretion process, the RN-dS black hole cannot always be overcharged and hence the WCCC is strongly respected beyond certain threshold limit. This result is always supported by non-linear order accretion, and so no violation of the CCC in the weak form occurs.

\section*{Acknowledgments}
The authors thank to the anonymous referees for
the valuable comments. S.S. and B.A. acknowledge the support of the Uzbekistan Ministry for Innovative Development.

\bibliographystyle{apsrev4-1}  
\bibliography{gravreferences}

 \end{document}